\documentstyle[aps,prb,epsf,psfig,multicol]{revtex}

\flushcolumns
\begin{document}
\draft

\widetext
\title{Microscopic conditions favoring itinerant ferromagnetism}
\author{J. Wahle,\cite{JWahle} N. Bl\"umer, J. Schlipf, K. Held,
and D. Vollhardt}

\address{Theoretische Physik III, Elektronische Korrelationen und
Magnetismus, Universit\"at Augsburg, D-86135 Augsburg, Germany}
\date{\today}
\maketitle

\begin{abstract}
  A systematic investigation of the microscopic conditions stabilizing
  itinerant ferromagnetism of correlated electrons in a single-band
  model is presented. Quantitative results are obtained by quantum
  Monte Carlo simulations for a model with Hubbard interaction $U$ and
  direct Heisenberg exchange interaction $F$ within the dynamical
  mean-field theory. Special emphasis is placed on the investigation
  of (i) the distribution of spectral weight in the density of states,
  (ii) the importance of genuine correlations, and (iii) the
  significance of the direct exchange, for the stability of itinerant
  ferromagnetism at finite temperatures. We find that already a
  moderately strong peak in the density of states near the band edge
  suffices to stabilize ferromagnetism at intermediate $U$-values in a
  broad range of electron densities $n$. Correlation effects prove to
  be essential: Slater--Hartree-Fock results for the transition
  temperature are both qualitatively and quantitatively incorrect. The
  nearest-neighbor Heisenberg exchange does not, in general, play a
  decisive role. Detailed results for the magnetic phase diagram as a
  function of $U$, $F$, $n$, temperature $T$, and the asymmetry of the
  density of states are presented and discussed.  
\end{abstract}

\begin{multicols}{2}\narrowtext

\section{Introduction}
\label{intro}
In contrast to conventional superconductivity and antiferromagnetism,
metallic ferromagnetism is in general an intermediate or strong
coupling phenomenon. Since there do not exist systematic investigation
schemes to solve such types of problems the stability of metallic
ferromagnetism is still not sufficiently understood. This is true even
within the simplest electronic correlation model, the one-band Hubbard
model,\cite{com1} in spite of significant progress made recently. The
Hubbard interaction is very unspecific, i.e., does not depend on the
lattice structure or dimension. Hence the
dispersion, and thereby the shape of the density of states (DOS), is of
considerable importance for the stability of ferromagnetism. This was
recognized already by Gutzwiller,\cite{Gutzwiller63}
Hubbard,\cite{Hubbard63} and Kanamori\cite{Kanamori63} in their
initial work on the Hubbard model.  However, the approximations used
in the early days of many-body theory were usually not reliable enough
to provide definite conclusions. An exception are the exact results by
Nagaoka\cite{Nagaoka66} on the stability of ferromagnetism at
$U=\infty$ in the case of one electron above or below half filling.
They show an important lattice sensitivity but, unfortunately, are not
applicable in the thermodynamic limit.

Over the years the stability of metallic ferromagnetism has turned out
to be a particularly difficult many-body problem whose explanation
requires subtle {\it nonperturbative} techniques. There has been an
upsurge of interest in this topic most recently
\cite{Lieb95,Mielke91,Penc96,MHgroup,Uhrig96,Daul97,Hlubina97,Ulmke98,Herrmann97ab,Vollhardt97b}.
These investigations confirm that ferromagnetism is favored in systems
with (i) frustrated lattices (which suppress antiferromagnetism) and
(ii) high spectral weight near the band edge closest to the Fermi
energy (which improve the kinetic energy of the polarized
electrons). Taken together, these properties imply a strongly
asymmetric DOS of the electrons.  Ferromagnetism on bipartite lattices
having a symmetric DOS may still be possible, but seems to require
very large values of $U$.\cite{Obermeier97} With the exception of
Refs.~\onlinecite{Ulmke98,Herrmann97ab} all previous
calculations refer to the ground state. It is therefore of interest to
obtain an answer to the question: {\it How does the distribution of
spectral weight in the DOS influence the stability regime of
ferromagnetism at finite temperatures?}

It should be noted that a strongly peaked, asymmetric DOS is a
considerably more complex condition for ferromagnetism than the Stoner
criterion. The latter merely asserts that, at $T=0$, the critical
interaction for the instability is determined by the inverse of the
DOS precisely {\it at} the Fermi energy $E_F$, $U_c=1/N(E_F)$, thus
neglecting antiferromagnetism and the structure of the DOS away from
$E_F$. Stoner (i.e., Hartree-Fock\cite{com5}) theory is a purely static
mean-field theory which ignores correlation effects,
e.g.,~the correlation-induced redistribution of momentum states and
the dynamic renormalizations of the band shape and width.
So the question remains:
{\it How essential are genuine correlation effects for the stability
  of itinerant ferromagnetism at finite temperatures?}

A third question concerns the suitability of the Hubbard model itself
as a model for ferromagnetism. Indeed there is no compelling {\it a
  priori} reason why the Hubbard model should be a good model for
ferromagnetism at all. Not only does it neglect band degeneracy, a
feature observed in all ferromagnetic transition metals (Fe, Co, Ni),
it also ignores the (weak) direct Heisenberg exchange interaction
which is equivalent to a ferromagnetic spin-spin interaction and hence
favors ferromagnetism in the most obvious
way.\cite{Gammeletc,Hirsch89a,Hirschetc,Strack95,Amadon96,Kollar96}
The proposition by Hirsch and
coworkers\cite{Hirsch89a,Hirschetc,Amadon96} that this interaction plays a
key role in metallic ferromagnetism was disputed by
Campbell et al.\cite{Gammeletc} So the controversial question is:

{\it How important is the direct Heisenberg exchange interaction for
  the stability of itinerant ferromagnetism in the one-band Hubbard
  model at finite temperatures?}

In this paper quantitative answers to the three questions formulated
above are given within the dynamical mean field theory (DMFT). The
DMFT, a nonperturbative approach, becomes exact in the limit of large
coordination
numbers.\cite{Metzner89a,MuellerHartmann89a,MuellerHartmann89b,Janisetc,Vollhardt93,Georges96}
When applied to $d=3$, where the coordination number is $O(10)$, the
DMFT has proven to yield accurate and reliable results, especially in
the context of long-range magnetic order.\cite{Rozenbergetc,Ulmke98}
It treats local correlations exactly while spatial
fluctuations are neglected. In this situation the momentum integral
entering in the local propagator will be replaced by an energy
integral involving only the DOS of the noninteracting electrons. The
latter may be viewed as an input parameter. In our investigation the
question concerning the importance of the distribution of spectral
weight within the band for the stability of ferromagnetism will
therefore be studied using a model DOS of the noninteracting electrons
whose shape can be changed continuously from symmetric to strongly
asymmetric by varying an asymmetry parameter.

The paper is structured as follows: in Sec.~II we present the model
under investigation, the dynamical mean-field equations, and the
analytical and numerical steps needed to construct magnetic phase
diagrams. The model DOS is introduced in Sec.~III. The results of our
investigation and quantitative answers to the questions posed above
are presented in Sec.~IV. A discussion where these results are put
into perspective (Sec.~V) closes the presentation.


\section{Model and methods of solution}
\label{QMC}

\subsection{Hubbard model with nearest-neighbor exchange}
The minimal model allowing one to treat an asymmetric
DOS, electronic on-site correlations, and the 
nearest-neighbor  Heisenberg
exchange interaction is given by
\begin{eqnarray}
  \hat{H} &=& \hat{H}_{\text{Hub}}- 2 F \sum_{\langle i,j
  \rangle }{\bf{\hat{S}}}_i\cdot{\bf{\hat{S}}}_j, 
\quad\mbox{where}\label{exthubm}\\
\hat{H}_{\text{Hub}} &=&  -\sum_{i,j,\sigma}  t_{ij}(\hat{c}^\dagger_{i  \sigma}
  \hat{c}^{\phantom{\dagger}}_{ j  \sigma} + {\rm h.c.}) + U \sum_{i }
  \hat{n}_{i\downarrow} \hat{n}_{ i  \uparrow}
\label{hub}
\end{eqnarray}
 Here ${\bf {\hat{S}}}_{i}=\frac{1}{2} \sum_{\sigma\sigma'}
\hat{c}^\dagger_{i\sigma} \bbox{\tau}_{\sigma\sigma'} \hat{c}^{\phantom
  \dagger}_{i\sigma'}$ with the vector of Pauli matrices $\bbox{\tau}$. 

We note that there are three other nearest-neighbor contributions
of the Coulomb interaction which might also effect the stability of the
ferromagnetic phase (see appendix).

\subsection{Dynamical mean-field theory}
Within the DMFT the coupling constants in (\ref{exthubm}) and
(\ref{hub}) have to be scaled with the lattice coordination number $Z$
as\cite{Metzner89a,MuellerHartmann89a} $t=t^*/\sqrt{Z}, F=F^*/Z$, 
where we consider nearest-neighbor hopping $t$
only. By analogy to classical spin models\cite{com2} the Hartree-Fock
approximation yields the exact result for the $F$-term in high
dimensions.

In the following we investigate the influence of the direct exchange
term on the properties of the Hubbard model in $d\to\infty$. Since the
Hubbard model is SU(2) spin-symmetric we can, without loss of
generality, assume a magnetization parallel to the $z$-axis. The
Hartree\cite{noteFock} decoupling then takes the form
${\bf{\hat{S}}}_i\cdot{\bf{\hat{S}}}_j \to \langle \hat{S}_i^z \rangle
\hat{S}_j^z + \hat{S}_i^z \langle \hat{S}_j^z \rangle - \langle
\hat{S}_i^z \rangle \langle \hat{S}_j^z \rangle$. In terms of the
magnetization $\hat{m}=\sum_i\hat{m}_i/N$ and its expectation value
$m=\langle \hat{m}\rangle$, where
$\hat{m}_i=2\,\hat{S}_i^z=\hat{n}_{i\uparrow}-\hat{n}_{i\downarrow}$
and $N$ is the number of lattice sites, the Hamiltonian
(\ref{exthubm}) can be written as\cite{noteAB}

\begin{equation}
\label{Hfnn}
  \hat{H} = \hat{H}_{\text{Hub}}- \frac{NF^*}{2} \,m\,\hat{m}
  +\frac{NF^*}{4} m^2.
\end{equation}
Apparently the influence of the exchange term in the limit $d\to
\infty$ is that of a (Weiss) magnetic field which vanishes
in the paramagnetic phase ($m=0$). Therefore, in this phase, all
one-particle properties of the system are those of the pure Hubbard
model. However, two-particle functions, especially the ferromagnetic
susceptibility, are modified (see Sec.~\ref{IIIB}).

In $d=\infty$, the Hubbard model (\ref{hub}) is equivalent to an
Anderson impurity model complemented by a self-consistency
condition.\cite{Jarrelletc,Georges96} Written in terms of Matsubara
frequencies $\omega_n$, self energy $\Sigma_\sigma$, the DOS of the
noninteracting electrons $N^0(\varepsilon)$, and thermal average
$\langle \psi \psi^* \rangle_{\cal A}$ the resulting coupled equations
for the Green function in the homogeneous phase read:
\begin{eqnarray}
G_\sigma(i\omega_n) &= &
\int\limits_{-\infty}^\infty d \varepsilon \,\frac{N^0 (\varepsilon)}
{i \omega_n +  \mu - \Sigma_\sigma(i\omega_n)- \varepsilon} \label{sk2}  \\
G_\sigma(i\omega_n) &= &
- \langle \psi_{\sigma n}^{\phantom *} \psi_{\sigma n}^*
\rangle_{\cal A}.\label{sk1}
\end{eqnarray}

The solution of the {\boldmath $k$}--integrated Dyson
equation~(\ref{sk2}) is straightforward and can be performed
analytically for the DOS used in this paper (see Sec.~IV). By contrast
the solution of Eq.~(\ref{sk1}) is highly nontrivial (for details of
the notation see Ref. \onlinecite{Ulmke95b}). It is achieved
using the auxiliary-field quantum Monte Carlo (QMC) algorithm by Hirsch and
Fye,\cite{Hirsch86} where a discretization of imaginary time,
$\Delta\tau=\beta/\Lambda$, is introduced. Here, $\Lambda$ denotes the
number of independent Matsubara frequencies. Physical quantities are
obtained in the $\Delta\tau\to 0$ limit.

\subsection{Calculation of susceptibilities, extrapolation and error
  handling} \label{IIIB} The second order phase transition from a
  para- to a ferromagnetic phase occurs at the zero of the inverse
  susceptibility $\chi_f^{-1}$, calculated\cite{Ulmke95b,Held97} in
  the paramagnetic phase. It is sufficient to perform all simulations
  for the pure Hubbard model since the influence of $F^*$ on the
  susceptibility is given by the following random-phase approximation
  (RPA)-like expression:
\begin{equation}
\label{RPA}
 \chi_f(U,F^*,\dots)=\frac{\chi_f(U,0,\dots)}{1-\frac{F^*}{2}
 \chi_f(U,0,\dots)},
\end{equation}
where $\chi_f(U,0,\dots)$ is the susceptibility of the pure Hubbard
model. This type of relation holds for pairs of two-particle
interactions (here $U$ and $F^*$) in arbitrary dimensions, whenever
one interaction (here $F^*$) is treated in Hartree-Fock approximation
and the other one (here $U$) exactly.\cite{com3}
In general, since the calculation of the susceptibility involves the
derivative of the self-energy $\Sigma_{U,F^*}$ with respect to some
field $h$,\cite{Ulmke95b,Held97} this follows from the fact that the
self-energy of the full Hamiltonian can be expressed as
\begin{equation}
\label{Sigma}
 \Sigma_{U,F^*}[G_{U,F^*}]=\Sigma_{U,0}[G_{U,F^*}]
 +\Sigma_{0,F^*}^{(1)}[G_{U,F^*}].
\end{equation}
Here $\Sigma[G]$ refers to the diagrammatic skeleton expansion of
$\Sigma$, where all lines are fully dressed propagators $G$. Since the
Hartree-Fock term $\Sigma_{0,F^*}^{(1)}$ only contributes in the
symmetry-broken phase, all $F^*$-renormalizations of $G_{U,F^*}$
vanish in the symmetric phase. Evaluating $d\Sigma_{U,F^*}/dh$ in the
symmetric phase, the first term in (\ref{Sigma}) leads to the same
contributions as without $F^*$-interaction, while the second term
introduces the RPA-like term proportional to $F^*$. For the $t$-$J$
model in DMFT the analogue of (\ref{RPA}) was derived by Pruschke et
al.\cite{Pruschke96}

Since $F^*$ only enters the calculations via Eq.~(\ref{Sigma}) we are
left with four physical parameters of the pure Hubbard model: Hubbard
interaction $U$, electron density $n$, temperature $T$ and an
asymmetry parameter $a$ for the kinetic energy (see Sec.~IV).
For each set of these five parameters Eqs.~(\ref{sk2}) and (\ref{sk1})
are iterated with typically $6\times 10^4$ Monte Carlo sweeps until
convergence is reached, i.e., the difference between two consecutive
values of ({\boldmath $\cal G$}$_\sigma)^{-1} = $
({\boldmath $G$}$_\sigma)^{-1} -$ {\boldmath $\Sigma$}$_\sigma$ 
is smaller than
$5\times 10^{-4}$ (measured by the norm $(2\Lambda)^{-1} \sum_{\sigma
n}|({\cal G}_{\sigma n}^{\mathrm{new}})^{-1} - ({\cal G}_{\sigma
n}^{\mathrm{old}})^{-1}|$; the energy scale is defined in
Sec.~IV). Subsequently eight measurements of the susceptibility are
performed with a reduced number of $2\times 10^4$ Monte Carlo
sweeps. Thus the result for each parameter set consists of an averaged
susceptibility $\chi_f(\Delta\tau)$ and its statistical error
$\Delta\chi_f(\Delta\tau)$. We neglect the propagation of the error in
({\boldmath $\cal G$}$_\sigma)^{-1}$ since it is always an order of
magnitude smaller than $\Delta\chi_f(\Delta\tau)$.
The extrapolation to $\Delta\tau = 0$ is performed by a quadratic
least squares fit of $\chi_f(\Delta\tau)$, using at least six
different values of $\chi_f$ for $\Delta\tau\in[0.09,0.5]$. 
Further details regarding the technical treatment can be found in
Refs.~\onlinecite{Ulmke95b,Ulmke95a}.

For mean-field theories like the DMFT a linear behavior of the inverse
susceptibility, i.e., a Curie-Weiss law, is expected and observed in
the vicinity of the transition. Thus the Curie temperature $T_C$ can
be obtained as the zero of a linear fit of $\chi_f^{-1}(T)$ drawn from
values of $\chi_f$ for four to six different temperatures (see,
e.g.,~Figs.~\ref{FvsT} and~\ref{FvsTa}, where $F^*_c=2\chi_f^{-1}$
(\ref{RPA}) is plotted).  The error of $T_C$ is obtained from the
errors $\Delta\chi_f(\Delta\tau)$ by error propagation and therefore
denotes only statistical, not systematic errors (e.g., due to the
extrapolation schemes used). However, we checked the accuracy of our
results by varying the procedure, e.g., extrapolating
$\chi_f^{-1}(\Delta\tau)$ instead of $\chi_f(\Delta\tau)$.


\section{Model spectral function}

Due to the vanishing of spatial fluctuations within the DMFT the
topology of the underlying lattice enters the self-consistency
equation~(\ref{sk2}) only via the noninteracting DOS, at least for
homogeneous phases. The choice of a particular model spectral function
thus represents a special (not unique) set of hopping elements
$t_{ij}$ in the Hamiltonian~(\ref{exthubm}) which characterize the
structure of the underlying lattice. Contributions to the kinetic
energy by, e.g., next-nearest-neighbor hopping can lead to an
asymmetrically shaped DOS, which is apparently favorable for the
stability of ferromagnetism. To investigate this stabilizing effect
quantitatively, we propose a model DOS with a shape-controlling
parameter. This parameter allows us to change smoothly from a
noninteracting DOS with (i) a symmetric shape (mimicking
nearest-neighbor hopping on a bipartite lattice) to (ii) an
asymmetrically peaked DOS (similar to a cubic lattice with
next-nearest-neighbor hopping) to (iii) a DOS with a square-root
divergence at the band edge (e.g., a fcc lattice with
next-nearest-neighbor hopping $t'=t/2$). The shape of the model DOS
thus qualitatively captures key features of real lattices.

\begin{figure}[t]
\def\epsfsize#1#2{\hsize}\epsfbox{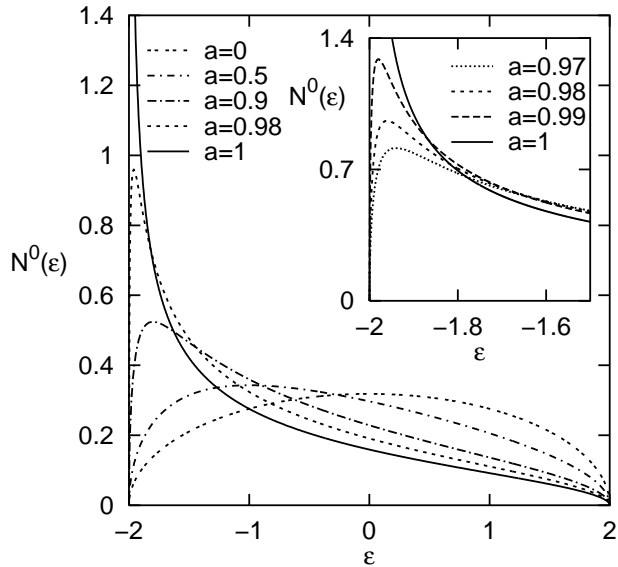}
\caption{Model spectral function (\ref{NDOS}) shown for different 
values of the asymmetry parameter $a$. By increasing $a$ spectral 
weight is shifted towards the lower band edge. The energy scale is
fixed by setting the variance of the DOS equal to 1.
\label{DOS}}
\end{figure}

The spectral function which we use throughout the paper is given by
\begin{equation}
N^0(\varepsilon)=c\,\frac{\sqrt{D^2-\varepsilon^2}}{D+a\varepsilon}
\label{NDOS}
\end{equation}
with $c=(1+\sqrt{1-a^2})/(\pi D)$ and bandwidth $2D$. The well-known
semielliptic DOS of the Bethe lattice with infinite number of nearest
neighbors is recovered for $a=0$. By increasing the parameter $a$
spectral weight is shifted towards the lower band edge
(Fig.~\ref{DOS}). For $a=1$ the DOS diverges at the lower band edge
like an inverse square-root. The particular choice of the model DOS
(\ref{NDOS}) has the advantage that the numerical effort of solving
the self-consistency equation (\ref{sk2}) is rather small since the
Hilbert transform can be calculated analytically. In the following we
set the variance to unity,
$\int\!d\varepsilon\,N^0(\varepsilon)\,\varepsilon^2-[\int\!d\varepsilon\,
N^0(\varepsilon)\,\varepsilon]^2=1$, thereby fixing the energy scale.
This leads to $D=2$ for all values of $a$. For $a=0$ it is equivalent
to choosing $t^*=1$ on the Bethe lattice.

While for the study of ferromagnetism within the DMFT the lattice
structure only enters via the DOS it is possible to construct
(infinitely many) corresponding dispersion relations
$\varepsilon$({\boldmath $k$}) or, equivalently, sets of hopping
elements $t_{i j}$. A realization in $d=1$ that is symmetric,
$\varepsilon (k) = \varepsilon(-k)$, and monotonous, $d \varepsilon /
d k > 0$, for $\varepsilon >0$ is given by
\begin{equation}
  \int\limits_{\varepsilon_{\text{min}}}^{\varepsilon} d\varepsilon' 
  N^0(\varepsilon')=\int\limits_0^k\frac{dk'}{\pi}=\frac{k(\varepsilon)}{\pi},
\end{equation}
where an inversion yields $\varepsilon(k)$. Generalizations to other
dimensions are possible with, e.g.,~$\varepsilon$({\boldmath $k$}$) =
\varepsilon(|${\boldmath $k$}$|)$\cite{Kollar97pc} or
$\varepsilon$({\boldmath $k$}$) = \sum_{i=1}^d
\varepsilon(k_i).$\cite{Ulmke97pc}

Although in principle one could thus choose a lattice corresponding to
the DOS~(\ref{NDOS}), this will not be done here, since we only study
homogeneous phases. Antiferromagnetism or incommensurate phases are
not expected to be important far away from half filling. Only for the
case $a=0$ (Bethe lattice) at $n=1$ the stability of an
antiferromagnetic phase is investigated.


\section{Results}

\subsection{The importance of the direct exchange interaction}
\begin{figure}[h]
\def\epsfsize#1#2{\hsize}\epsfbox{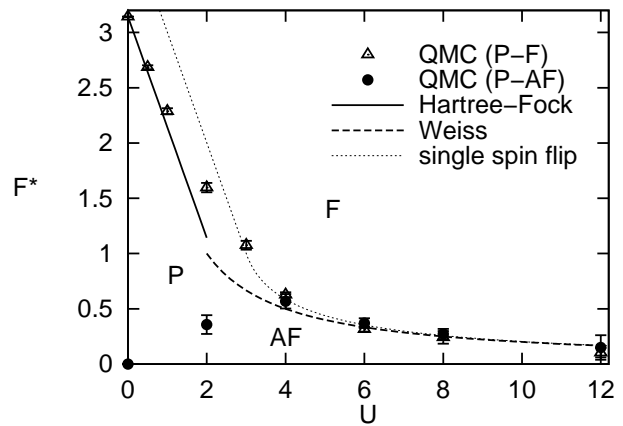}
\caption{Phase diagram for a symmetric DOS ($a=0$):~direct exchange 
coupling $F^*$ vs.~Hubbard 
interaction $U$ at half filling ($n=1$) extrapolated to $T=0$. Open 
triangles and filled circles correspond to the instability of the 
paramagnetic phase (P) against the ferromagnetic (F) and 
antiferromagnetic (AF) order, respectively. Solid line: Hartree-Fock 
theory; dashed line: Weiss mean-field theory; dotted line:
sinle spin flip instability for the saturated ferromagnetic state.}
\label{Fsymm}
\end{figure}

On a bipartite lattice with perfect nesting, the ground state of the
pure Hubbard model at half filling is antiferromagnetic for all $U>0$,
at least in dimensions $d\ge 3$. In this situation a ferromagnetic
state is strongly disfavored also in the general model,
Eq.~(\ref{exthubm}). At large $U$, however, when the model reduces to
an effective Heisenberg model (which, in high dimensions, is exactly
described by Weiss mean-field theory), already a small value of the
direct exchange interaction, $F^*> 2 (t^*)^2/U$, is sufficient to
stabilize a ferromagnetic ground
state.\cite{Campbell88,Tang90,Strack95} Indeed, the Heisenberg model
well describes the $F^*$--$U$ ground state phase diagram at half
filling down to $U\approx 4$; this is evident from Fig.~\ref{Fsymm},
where a comparison with our QMC results is shown for a symmetric DOS
[Eq.~(\ref{NDOS}) with $a=0$]. At small $U$ the phase boundary between
a paramagnetic and a ferromagnetic state is correctly reproduced by
Hartree-Fock theory (but only for $U< 1$). This is not surprising
since in $d=\infty$ the $F$-term is treated exactly within this
approximation. Also included in Fig.~\ref{Fsymm} is the line below
which a fully saturated ferromagnetic state becomes unstable against
single spin flips as first computed by Hirsch\cite{Hirsch90b} for cubic
lattices. For the Bethe lattice this line is given exactly by the
Hartree-Fock result $F^* = 4-U$ for $U\le U_c=3$ and
\begin{equation}
F^*-U = - \frac{8}{(F^*+U)\left(1-\sqrt{1-16/(F^*+U)^2}\right)}
\end{equation}
for $U>3$. This can be seen from
Eqs. 5 and 7 in Ref.~\onlinecite{Hirsch90b} and the known analytic
expression for the Hilbert transform of the semielliptic DOS. The
remarkable agreement between the QMC results and this curve for $U\ge 3$ 
suggests that (at zero temperature) the region of partial
polarization is very narrow already at intermediate interaction
strength $U$.

\begin{figure}[ht]
\def\epsfsize#1#2{\hsize}\epsfbox{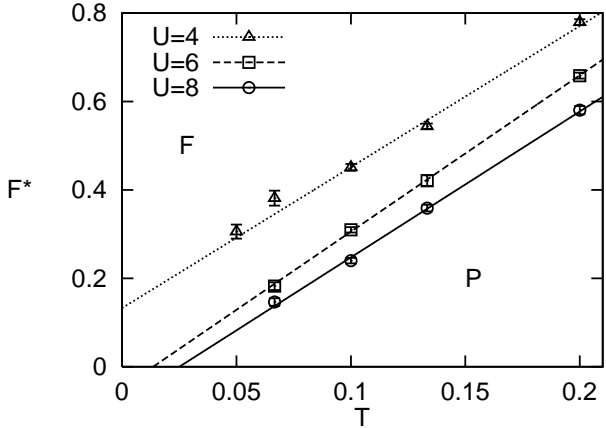}
\caption{$F^*$--$T$ phase diagram for different values of $U$ for a
strongly peaked DOS ($a=0.98$) at a filling $n=0.6$. The linear
extrapolation shows that there exists a critical $U_c$ above
which ferromagnetism is stable even without the direct exchange
coupling.}
\label{FvsT}
\end{figure}

The $F^*$--$T$ phase diagram for a strongly peaked DOS ($a=0.98$) at
filling $n=0.6$ is shown in Fig.~\ref{FvsT}. The QMC results for the
ferromagnetic phase boundary can be extrapolated linearly to zero
temperature leading to a ground state phase diagram. Clearly the
values of $F^*$ necessary to stabilize ferromagnetism are
significantly reduced in comparison to the bipartite case. In
particular, for $U=6$ and $U=8$ the extrapolation lines cross the
ordinate at positive temperatures. Thus, an asymmetric DOS 
\hfill{\it stabilizes \hfill ferromagnetism \hfill even \hfill in
\hfill the \hfill pure \hfill Hubbard} 
\noindent
{\it model} ($F^*=0$)
above a critical interaction strength $U_c$ with $4< U_c < 6$.

\begin{figure}[ht]
\def\epsfsize#1#2{\hsize}\epsfbox{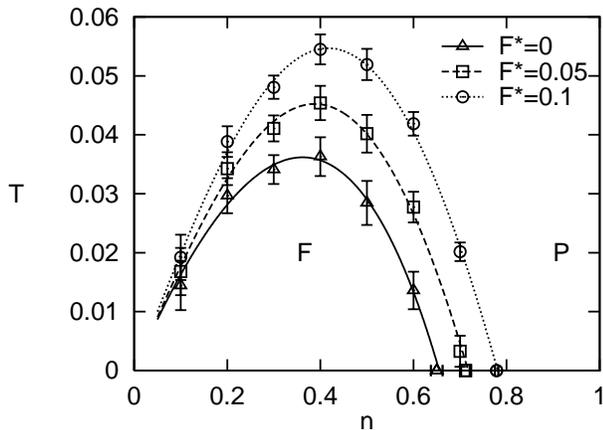}
\caption{$T$--$n$ phase diagram for different values of the direct 
exchange coupling $F^*$ for a strongly peaked DOS ($a=0.98$) at $U=6$.
A small direct exchange coupling is seen to enlarge the stability 
regime of the ferromagnet, especially for densities close to 
half filling. The lines are guide to the eye only.}
\label{TvsnF}
\end{figure}

Figs.~\ref{FvsT} and~\ref{TvsnF} prove that already a small direct
exchange coupling $F^*$ can significantly enhance ferromagnetic
tendencies and thus give the final ``kick'' towards ferromagnetism for
systems that are close to an instability. This influence is stronger
at larger densities when the local magnetic moments are enhanced
(Fig.~\ref{TvsnF}). The lower critical densities are very small, but
larger than those predicted by Hartree-Fock theory, since Hartree-Fock
always overestimates the size of the ferromagnetic regime.

\vspace{1cm}

Quite generally, the value of the exchange interaction $F$ in metals
can be expected to be rather small. Note, however, that for
three-dimensional lattices the {\it scaled} quantity $F^*=Z F$ is an order
of magnitude larger than the exchange coupling $F$ itself. Hubbard's
crude estimate\cite{Hubbard63} of $F/U \approx 1/400$ would therefore
imply that, e.g., at $U=6$ a (scaled) exchange interaction as large
as $F^*=0.15$ is not completely unrealistic.

\subsection{The importance of the asymmetry of the DOS}
\begin{figure}[ht]
\def\epsfsize#1#2{\hsize}\epsfbox{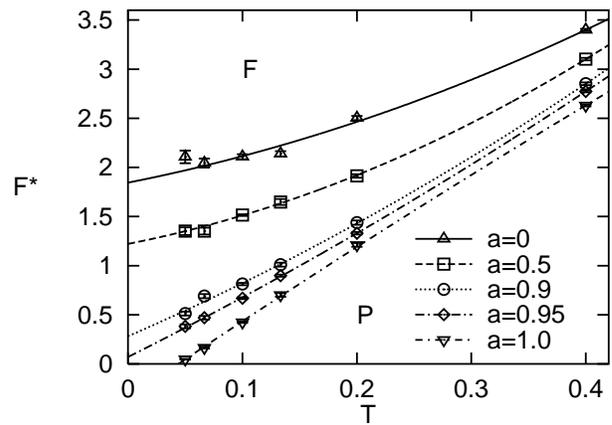}
\caption{$F^*$--$T$ phase diagram for different shapes of the DOS at 
$U=4$ and $n=0.3$. For values of $0.95<a\le 1$ the ferromagnetic phase 
is stable even without the direct exchange coupling. The lines show 
a quadratic least-squares fit in $T$.}
\label{FvsTa}
\end{figure}

The dependence of the phase boundary on the asymmetry parameter $a$ is
systematically studied in Fig.~\ref{FvsTa} at $U=4$ for a relatively
small electron density $n=0.3$. For a symmetric or slightly asymmetric
DOS ($a=0$, $a=0.5$) the system only becomes ferromagnetic for $F^*>1$
even at $T=0$. For $a=0.9$, when the shape of the DOS is roughly
triangular, the critical $F^*$ is considerably reduced. But only when
a marked peak develops (i.e., for $a>0.95$) does the critical $F^*$
drop to zero; ferromagnetism is then stable even in the pure Hubbard
model. From now on we restrict our studies to this case ($F^*=0$).

\begin{figure}[ht]
\def\epsfsize#1#2{\hsize}\epsfbox{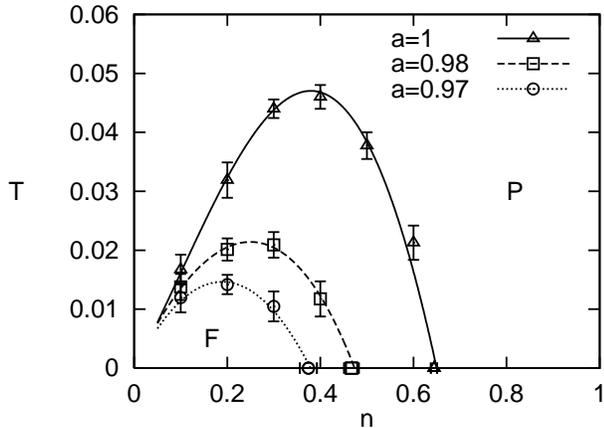}
\caption{$T$--$n$ phase diagram for different shapes of the DOS at 
$U=4$. By shifting spectral weight towards the lower band edge, the
region of stability of the ferromagnetic phase is enlarged. The lines
are guide to the eye only.}
\label{Tvsna}
\end{figure}

The $T$ vs.~$n$ phase diagram is shown in Fig.~\ref{Tvsna} for $U=4$
and three different shapes of the DOS ranging from strongly asymmetric
($a=0.97$) to divergent at the lower band edge ($a=1$). Evidently the
ferromagnetic phase is largest at $a=1$. We want to stress however,
that the divergence does {\em not} change the physics qualitatively
(except at $n\ll 1$). A moderately strong peak near the band edge is
all that is needed to stabilize ferromagnetism.

\vspace{1cm}

For symmetric densities of states at $U<12$ (Bethe
lattice and hypercubic lattice) we find {\it suppression} of the
ferromagnetic (as well as the antiferromagnetic) susceptibility away
from half filling.\cite{Bluemer96} This does not exclude the
possibility for ferromagnetism on bipartite lattices at much larger
values of $U$. Indeed, very recently ferromagnetism was found within
the noncrossing approximation for the $t$--$J$ model on a hypercubic
lattice in the limit $d\to \infty$ for $U>30$ away from half
filling.\cite{Obermeier97} Apparently, at least for moderate $U$, the
bipartite lattice with only nearest-neighbor-hopping is not a natural
``environment'' for ferromagnetism -- the asymmetry of the
noninteracting DOS is crucial.

\subsection{The importance of correlations}
\begin{figure}[ht]
\def\epsfsize#1#2{\hsize}\epsfbox{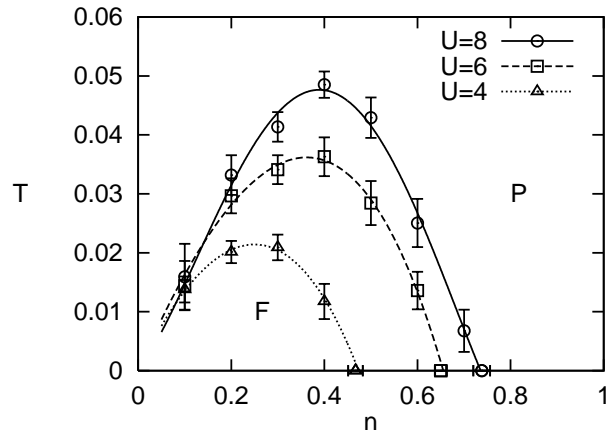}
\caption{$T$--$n$ phase diagram for a strongly peaked DOS ($a=0.98$) 
for different values of $U$. With increasing Hubbard interaction the 
stability regime of the ferromagnetic phase becomes larger, especially 
at higher densities. The lines are guide to the eye only.}
\label{TvsnU}
\end{figure}

In Fig.~\ref{TvsnU} the $T$--$n$ phase diagram is shown for different
values of the on-site interaction $U$. Evidently the ferromagnetic
phase becomes more favorable for increasing $U$: both the maximal
Curie temperature and the (upper) critical density rise. This effect
is seen to be qualitatively similar to an increase of the exchange
interaction $F^*$ or the asymmetry of the DOS $a$ (Figs.~\ref{TvsnF}
and~\ref{Tvsna}, respectively).

\begin{figure}[ht]
\def\epsfsize#1#2{\hsize}\epsfbox{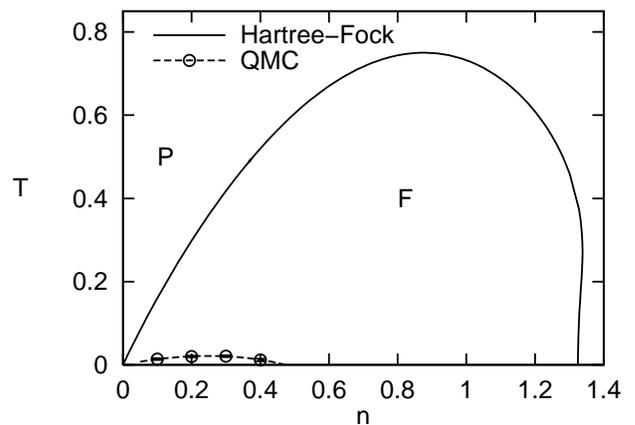}
\caption{$T$--$n$ phase diagram for a strongly peaked DOS ($a=0.98$) 
at $U=4$. The comparison between Hartree-Fock theory (solid line) 
and DMFT (QMC, circles) reveals the importance of correlation 
effects. The dashed line is guide to the eye only.}
\label{TvsnHF}
\end{figure}

\begin{figure}[ht]
\def\epsfsize#1#2{\hsize}\epsfbox{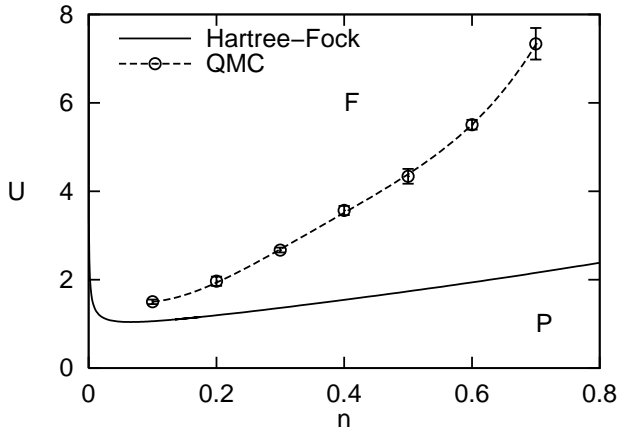}
\caption{$U$--$n$ phase diagram at $T=0$ for a strongly peaked DOS
($a=0.98$). The DMFT data (QMC, circles) is extrapolated from finite
temperature calculations. The Stoner criterion (solid line) 
underestimates the critical $U_c(n)$ for ferromagnetism, but becomes 
better at lower densities. The dashed line is guide to the eye only.}
\label{UvsnHF}
\end{figure}

\begin{figure}[ht]
\def\epsfsize#1#2{\hsize}\epsfbox{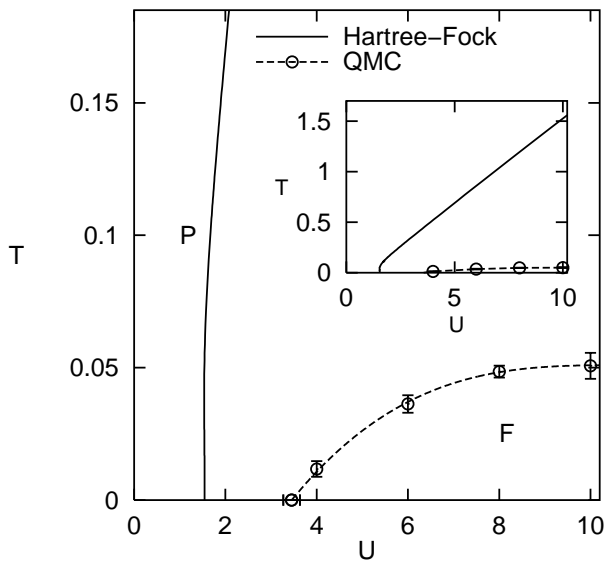}
\caption{$T$--$U$ phase diagram for a strongly peaked DOS ($a=0.98$) 
at $n=0.4$. The comparison between Hartree-Fock theory (solid line) and 
 DMFT (QMC, circles) shows the former can describe the Curie 
temperature $T_C(U)$ neither quantitatively, nor qualitatively. The 
dashed line is guide to the eye only. }
\label{TvsUHF}
\end{figure}

Our QMC results are compared with Hartree-Fock theory in
Figs.~\ref{TvsnHF}--\ref{TvsUHF}. We note that, applied to the Hubbard
model, the DMFT includes Hartree-Fock theory as its static limit and
is thus superior in any dimension. Fig.~\ref{TvsnHF} shows the vast
overestimation of the ferromagnetic phase within Hartree-Fock. The
maximal Curie temperature obtained in this approximation is more than
an order of magnitude too large. At such high temperatures details of
the DOS are averaged out and consequently the density dependence
(e.g., the position of the maximum) is completely artificial. In
Fig.~\ref{UvsnHF} the $U$ vs.~$n$ ground state phase diagram is shown
and compared to the Stoner criterion.\cite{com5} At low $n$ the Stoner
curve clearly approaches the QMC curve. Since the DOS vanishes
smoothly at the lower band edge for $a<1$ both curves diverge for $n
\to 0$. Figure~\ref{TvsUHF} focuses on the limit of large $U$. The
weak coupling Hartree-Fock theory fails again: it predicts an
unbounded linear increase of $T_C$ with $U$, $T_C \sim U n(2-n)/4$,
whereas QMC shows that $T_C$ has a finite limit for $U\to\infty$. It
is expected that such a finite limit exists for all densities. A
saturation is also suggested by the curves in Fig.~\ref{TvsnU}. It
arises from the suppression of double occupancies by correlations. In
contrast to the Hartree-Fock prediction the interaction energy goes to
zero for $U\to\infty$, thus only the bandwidth remains as an energy
scale. In the special case $a=0.98$, $n=0.4$ one finds\cite{com4}
$T_C(U=\infty)=0.07\pm 0.02$.

\vfill

One might argue that comparison of the DMFT results should not be made
with Hartree-Fock itself but with Hartree-Fock plus quantum
corrections, since the latter are known to reduce many of the
deficiencies of Hartree-Fock. Such corrections have been discussed by
van Dongen\cite{vanDongen91a} and Freericks and
Jarrell.\cite{Freericks95a} The latter authors showed how quantum
fluctuations modify the Stoner criterion by subtracting the
particle-particle susceptibility.  Evaluating these corrections in the
case of Fig.~\ref{TvsUHF} we find that the ferromagnetic phase is
completely suppressed (as previously observed in
Ref.~\onlinecite{Freericks95a} for a symmetric DOS). At $a=0.98$ this
holds for all densities $n\gtrsim 0.3$. Thus the ``corrections'' to
Hartree-Fock are seen to {\it underestimate} the ferromagnetic region
by far.


\section{Discussion and Outlook}
 
After more than three decades of research it has become clear at
last\cite{Daul97,Hlubina97,Ulmke98} that the Hubbard model can
describe itinerant ferromagnetism even on regular lattices and at
moderate $U$-values for a wide range of electronic densities $n$. Since
ferromagnetism is an intermediate to strong coupling problem the
question concerning its ``mechanism'' has, in principle, no
straightforward answer. This is in contrast to {\em weak} coupling
phenomena, e.g., conventional superconductivity, which can be
explained within perturbation theory. Nevertheless a good starting
point for an understanding of the origin of itinerant ferromagnetism
can be obtained in the strong coupling limit. At $U=\infty$ doubly
occupied sites are excluded and the Hubbard model reduces to a
(complicated) kinetic energy. To avoid doubly occupied sites in a
paramagnetic phase the DOS is then necessarily strongly renormalized
compared with the noninteracting case, whereas for the saturated
ferromagnetic phase the interacting DOS is the same as the
noninteracting one except for a shift between the spin-up and -down
bands. In this situation details of the structure of the
noninteracting DOS become relevant in selecting the state with the
lowest energy. This physical picture is, in principle, similar to that
underlying the Nagaoka mechanism; however, the latter only addresses
the kinetic energy of a single hole and it was so far not possible to
generalize it to thermodynamically relevant densities.
Our investigations within the DMFT explicitly show that a moderately
strong peak at the band edge closest to the Fermi energy is sufficient
to stabilize ferromagnetism.  Furthermore a strong asymmetry of the
DOS implies a nonbipartite lattice which frustrates the competing
antiferromagnetism near half filling.

The mechanism described above is completely different from the mere
band shift of the Hartree-Fock theory. This weak coupling approach
does not take into account the dynamical renormalization of the DOS in
the paramagnetic phase and thus predicts ferromagnetism for any DOS,
even at relatively small values of $U$ and for high temperatures.  The
comparison with DMFT clearly shows that Hartree-Fock (i) overestimates
transition temperatures by more than an order of magnitude, (ii)
renders the dependence of $T_C$ on $U$ qualitatively incorrect, and
(iii) predicts ferromagnetism for the symmetric DOS, where (at least
for $U<12$) this is not found. These shortcomings of Hartree-Fock
theory are due to the neglect of dynamical fluctuations which are at
the heart of the correlation problem.

The Heisenberg exchange interaction, not considered in the pure
Hubbard model, provides another mechanism that may order the
fluctuating local moments arising by the suppression of double
occupancies. We found that for a symmetric DOS rather large values of
$F$ are needed to stabilize ferromagnetism. However, for an asymmetric
DOS with a peak near the band edge already small values of the
exchange interaction may provide the final ``kick'' towards
ferromagnetism. In any case it reduces the critical on-site
interaction and increases the critical temperatures of the
ferromagnetic phase boundary.

While the DMFT correctly describes the dynamic fluctuations of the
interacting many-body system, it neglects spatial fluctuations and
short-range order. Hence one should suspect that this approach
overestimates the transition temperatures $T_C$. Within DMFT
Ulmke\cite{Ulmke98} estimated $T_C$ for a three-dimensional fcc
lattice to be of the order of 500 to 800 K which is in the range of
realistic transition temperatures. We may expect spatial fluctuations
to reduce these temperatures. On the other hand, band degeneracy, not
considered in our model so far, is expected to increase $T_C$. Indeed,
band degeneracy and Hund's rule couplings which are clearly present in
realistic systems can be rigorously shown to improve the stability of
ferromagnetism at least for special parameter
values.\cite{Kollar98,Vollhardt97b} The incorporation of band
degeneracy, for which the DMFT also provides a suitable framework, is
the most important feature that has to be included in future
investigations of the Hubbard model.\cite{Held98} The additional
nearest-neighbor interactions discussed in the appendix may provide yet
another mechanism for ferromagnetism and will be studied in the
future.

\subsection*{Acknowledgments}
We are very grateful to M.~Kollar and M.~Ulmke for numerous helpful
discussions. One of us (N.B.) thanks the Fulbright Commission for
support and the Department of Physics at the University of Illinois
for hospitality. Computations were performed on the Cray T90
of the HLRZ J\"ulich.


\begin{appendix}
\section*{Nearest-Neighbor Interactions}

In Wannier representation the Coulomb interaction gives rise
to the purely local interaction $U$ as well as to four nearest-neighbor
interactions. \cite{Gammeletc,Hirsch89a,Hirschetc,Vollhardt97b}
Besides the Heisenberg exchange interaction these are the
density-density interaction, the pair-hopping term, and
the off-diagonal ``bond-charge--site-charge''\cite{Gammeletc} interaction. 
The latter effectively describes a
density-dependent hopping \cite{Hirschetc} and leads to a narrowing
of the band. Hence this term 
is expected to stabilize saturated ferromagnetism. Since
the quantum dynamics of this term make a systematic investigation
difficult -- even within the DMFT -- its detailed study has to be postponed to
the future.
The pair-hopping term also weakly
enhances ferromagnetic tendencies.\cite{Hirschetc,Hirsch91}

Among all nearest-neighbor interactions the density-density term
\begin {equation}
\hat{H}^V_{\text{NN}} = 
V \sum_{\langle i,j\rangle} 
\hat{n}_i \hat {n}_j,
\end {equation}
is largest and is thus
investigated explicitly in the following. 
In the case of d-electrons Hubbard roughly estimated this
term to $V$=2-3 eV, an order of magnitude smaller than the Hubbard
interaction $U$.\cite{Hubbard63} However, since there are $Z$
neighbors contributing, the total energy of the nearest-neighbor
density-density interactions may in some materials even surpass that
of the Hubbard interaction. This raises the question of the importance
of the $V$-term, in particular its influence on the ferromagnetic
phases investigated in the present paper.

It was already pointed out by
M\"uller-Hartmann\cite{MuellerHartmann89a} that in the limit $d
\rightarrow \infty$ and with the proper scaling $V=V^*/Z$ the
nearest-neighbor density-density interaction reduces to its Hartree
contribution which may then be viewed as a simple, site-dependent
shift of the chemical potential. In the absence of broken
translational symmetry the chemical potential must compensate this
shift to keep the electron density fixed. Then there is no effect at
all.

On bipartite lattices translational symmetry can be broken by a charge
density wave (CDW) with different electron densities on A and B
sublattices, i.e., with order parameter $n_{CDW}=(n_A-n_B)/2$. To
study this possible ordering we analyze the instability towards a CDW
in the following.

Similar to the exchange term $F$ the Hartree contribution of the
interaction $V$ leads, even in the presence of other interactions, to
an RPA-like pole in the CDW susceptibility [cf.~Eq.~(\ref{RPA})]:\cite{com3}
\begin{displaymath}
\chi_{CDW}(U,V^*,\ldots) = \frac{\chi_{CDW}(U,0,\ldots)}
 {1-V^* \; \chi_{CDW}(U,0,\ldots)}\,.
\end{displaymath}
Thus a second order phase transition to a CDW occurs at
$V^*_c=1/\chi_{CDW}(U,0,\ldots)$.

\begin{figure}[ht]
\def\epsfsize#1#2{\hsize}\epsfbox{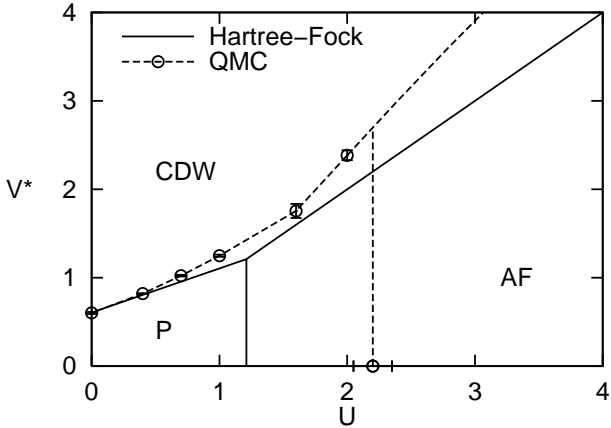}
\caption{$V^*$--$U$ phase diagram for the semielliptic DOS ($a=0$) at 
$T=0.125$. For $V^*>U$ a regime with charge density wave order (CDW) 
is established.
The antiferromagnetic phase vanishes for $U>7.7\pm0.5$ (not shown). The
dashed line is guide to the eye only.\label{VvsU} }
\end{figure}

Since next-nearest-neighbor hopping frustrates CDW order, the maximal
instability towards a CDW is expected for the symmetric DOS with $a=0$
in Eq.~(\ref{NDOS}). Half filling is optimal in this case. For these
parameters we determined the phase diagram Fig.~\ref{VvsU} employing
the QMC technique (for details concerning the calculation of the CDW
susceptibility see Ref.~\onlinecite{Held96}).
Within DMFT a CDW ordering occurs for
$V^*\gtrsim U$ (at not too high temperatures). Compared to the
Hartree-Fock approximation the CDW phase boundary of the full model is
only slightly moved towards larger values of $V^*$. A similar
deviation from Hartree-Fock was found by means of QMC simulations in
$d=1$ by Hirsch,\cite{Hirsch83b} in $d=2$ by Zhang 
and Callaway,\cite{Zhang89} and within perturbation theory for both
\noindent
weak and strong coupling (in arbitrary dimensions) by van
Dongen.\cite{vanDongen91a,vanDongen}

All these studies demonstrate that the CDW is stable for $V^*=V
Z>U$. While this relation may in principle hold for some transition
metals, ferromagnets apparently do not show spatial charge
ordering. Therefore the adequate correlated electron model for a
ferromagnet appears to have parameters in the range $V^* \le U$. Then
the nearest-neighbor density-density interaction $V^*$ has no
influence on the phase diagram, especially on the border of the
ferromagnetic phase, at least in $d=\infty$. Even in $d=3$ the Hartree
diagram gives the main contribution of the interaction $V^*$ since
spatial fluctuations, leading to genuine correlations, are suppressed
as $1/Z$. Moreover in $d=1$ and at half-filling the effect of $V^*$
on the ferromagnetic phase boundary is still small.\cite{Amadon96}
Therefore, over an extended range of parameters the
nearest-neighbor term $V^*$ has almost no influence and thus its
importance is seen to be much smaller than its value suggests.
\end{appendix}


\begin{thebibliography}{10}
\bibitem{com5}
For ferromagnetism in the Hubbard model, Hartree-Fock theory is equivalent to
  Stoner theory.
\bibitem{noteFock}
The Fock contribution of the $F$ term vanishes as $1/Z$ since it is given by
  two Green functions connecting neighboring sites. In $d=3$ the Fock term
  renormalizes the band-width: $\tilde{D}(E(\epsilon)) = D(\epsilon)/\{1-F^*/(Z
  {t^*}^2)\int d \epsilon D(\epsilon) (-2 \epsilon) f[E(\epsilon)]\}$ (see
  Eq.~(29) in Ref.\onlinecite{Hirsch89a}, which readily yields this equation
  for the paramagnetic phase). The renormalization is maximal at half-filling
  and reduces the critical value of $F^*$ for the onset of ferromagnetism,
  e.g., for the half-filled cubic lattice from $F^*=1.17 t$ to $F^*=0.84 t$
  within the Hartree-Fock approximation ($F^* (\tilde{D}(\epsilon_F)=1$). This
  effect of order $1/Z$ is neglected in the DMFT.
\bibitem{noteAB}
The definition of generalized (e.g., staggered) susceptibilities for
bipartite ($AB$-) lattices requires the introduction of an index
$\alpha \in\{A,B\}$ in Eq.~(\ref{Hfnn}) that indicates the
sublattice. The corresponding generalizations are straightforward
(see, e.g., Ref.~\onlinecite{Held97}).
\bibitem{com4}
The error mainly reflects an uncertainty in the extrapolation law. It was
  determined by comparing linear and quadratic least-squares fits in $1/U$.
\end{thebibliography}

\end{multicols}
\end{document}